\documentclass[journal,twoside,web]{ieeecolor}
\usepackage{jsen}
\usepackage{cite}
\usepackage{amsmath,amssymb,amsfonts}
\usepackage{algorithmic}
\usepackage{graphicx}
\usepackage{textcomp}
\usepackage{wrapfig}
\usepackage{amsmath}
\usepackage{ragged2e}
\usepackage{comment}
\usepackage{url}
\usepackage{float}
\usepackage{array}
\usepackage{multirow}
\usepackage{makecell}
\usepackage{tabularx}
\usepackage{booktabs}
\usepackage[acronym]{glossaries}
\usepackage{acronym}
\usepackage{multirow}   
\usepackage{makecell} 
\usepackage{siunitx}
\usepackage[flushleft]{threeparttable}
\usepackage{subcaption}
\usepackage{comment}
\usepackage{textcomp}

\newcommand{\review}[1]{{\textcolor{black}{#1}}}

\def\BibTeX{{\rm B\kern-.05em{\sc i\kern-.025em b}\kern-.08em
    T\kern-.1667em\lower.7ex\hbox{E}\kern-.125emX}}
\definecolor{abstractbg}{rgb}{0.89804,0.94510,0.83137}
\begin{document}
\acrodef{AW}{alpha waves}
\acrodef{PULP}{Parallel Ultra Low Power}
\acrodef{SoC}{System on Chip}
\acrodef{BLE}{Bluetooth Low Energy}
\acrodef{EEG}{electroencephalography}
\acrodef{EMG}{electromyography}
\acrodef{ECG}{electrocardiogram}
\acrodef{PPG}{photoplethysmogram}
\acrodef{EOG}{electrooculography}
\acrodef{MM}{Motor Movement}
\acrodef{SSVEP}{Steady State Visually Evoked Potential}
\acrodef{HMI}{human-machine interface}
\acrodef{PMIC}{Power Management Integrated Circuit}
\acrodef{IMU}{inertial measurement unit}
\acrodef{DSP}{digital signal processing}
\acrodef{NN}{neural network}
\acrodef{ULP}{ultra-low-power}
\acrodef{sEMG}{surface electromyography}
\acrodef{DL}{Deep Learning}
\acrodef{ML}{Machine Learning}
\acrodef{LOSO CV}{leave-one-subject-out cross-validation}
\acrodef{CV}{cross-validation}
\acrodef{BCI}{Brain-Computer Interface}
\acrodef{CCA}{canonical-correlation analysis}
\acrodef{SoA}{State-of-the-Art}
\acrodef{SNR}{signal-to-noise ratio}
\acrodef{PGA}{programmable-gain amplifier}
\acrodef{BTE}{Behind-the-Ear}
\acrodef{CNN}{Convolutional Neural Network}
\acrodef{DNN}{Deep Neural Network}
\acrodef{ITR}{information transfer rate}
\acrodef{NE16}{Neural Engine 16}
\acrodef{PTH}{plated through hole}
\acrodef{CMRR}{common-mode rejection ratio}
\acrodef{AFE}{analog frontend}
\acrodef{PCB}{printed circuit board}
\acrodef{IFCN}{International Federation of Clinical Neurophysiology}
\acrodef{NCCA}{Normalized Canonical Correlation Analysis}
\acrodef{ADC}{Analog to Digital Converter}
\acrodef{CV}{Cross Validation}
\acrodef{SoA}{State-of-the-Art}
\acrodef{SIMO}{Single-Inductor Multiple-Output}
\acrodef{LDO}{Low Drop-Out}
\acrodef{PSRAM}{Pseudo Static Random Access Memory}
\acrodef{WLCSP}{Wafer Level Chip Scale Package}
\acrodef{PDM}{Pulse Density Modulation}
\acrodef{MLC}{Machine Learning Core}
\acrodef{AAD}{Acoustic Activity Detect}
\acrodef{ASC}{Adaptive Self-Configuration}
\acrodef{RNN}{Recurrent Neural Network}
\acrodef{SFU}{Smart Filtering Unit}
\acrodef{SDK}{Software Development Kit}
\acrodef{MCU}{Micro-Controller Unit}
\acrodef{SoC}{System on Chip}
\acrodef{JTAG}{Joint Test Action Group}
\acrodef{GPIO}{General Purpose Input-Output}
\acrodef{RMS}{Root Mean Square}
\acrodef{PTT}{Pulse Transit Time}
\acrodef{PAT}{Pulse Arrival Time}
\acrodef{IOT}{Internet of the Things}
\acrodef{AI}{Airtificial Inteligence}
\acrodef{BPM}{Blood Pressure Monitoring}

\acrodef{SSVEP}{Steady-state Visual Evoked Potential}
\acrodef{ASSR}{Auditory Steady-State Response}
\acrodef{SoC}{System-on-Chip}
\acrodef{Resp}{Respiration}
\acrodef{Temp}{Temperature}
\acrodef{DD}{Drowsiness Detection}
\acrodef{FMG}{Force Myography}

\acrodef{OFA}{Once-For-All}
\acrodef{SIMD}{Single Instruction, Multiple Data}
\acrodef{ELU}{Exponential Linear Unit}
\acrodef{ReLU}{Rectified Linear Unit}
\acrodef{RPR}{Random Partition Relaxation}
\acrodef{MAC}{Multiply Accumulate}
\acrodef{DMA}{Direct Memory Access}
\acrodef{BMI}{Brain--Machine Interface}
\acrodef{BCI}{Brain--Computer Interface}
\acrodef{SMR}{Sensory Motor Rythms}
\acrodef{EEG}{Electroencephalography}
\acrodef{SVM}{Support Vector Machine}
\acrodef{SVD}{Singular Value Decomposition}
\acrodef{EVD}{Eigendecomposition}
\acrodef{IIR}{Infinite Impulse Response}
\acrodef{FIR}{Finite Impulse Response}
\acrodef{FC}{Fabric Controller}
\acrodef{NN}{Neural Network}
\acrodef{MRC}{Multiscale Riemannian Classifier}
\acrodef{FLOP}{Floating Point Operation}
\acrodef{SOS}{Second-Order Section}
\acrodef{IPC}{Instructions per Cycle}
\acrodef{TCDM}{Tightly Coupled Data Memory}
\acrodef{FPU}{Floating Point Unit}
\acrodef{FMA}{Fused Multiply Add}
\acrodef{ALU}{Arithmetic Logic Unit}
\acrodef{DSP}{Digital Signal Processing}
\acrodef{GPU}{Graphics Processing Unit}
\acrodef{SoC}{System-on-Chip}
\acrodef{MI}{Motor-Imagery}
\acrodef{CSP}{Commmon Spatial Patterns}
\acrodef{PULP}{parallel ultra-low power}
\acrodef{SoA}{state-of-the-art}
\acrodef{BN}{Batch Normalization}
\acrodef{ISA}{Instruction Set Architecture}
\acrodef{ECG}{Electrocardiogram}
\acrodef{RNN}{recurrent neural network}
\acrodef{CNN}{convolutional neural network}
\acrodef{TCN}{temporal convolutional network}
\acrodef{EMU}{epilepsy monitoring unit}
\acrodef{ML}{Machine Learning}
\acrodef{DL}{Deep Learning}
\acrodef{AI}{Artificial Intelligence}
\acrodef{TCP}{Temporal Central Parasagittal}
\acrodef{LOOCV}{Leave-One-Out Cross-Validation}
\acrodef{WFCV}{Walk-Forward Cross-Validation}
\acrodef{RWCV}{Rolling Window Cross-Validation}
\acrodef{IoT}{Internet of Things}
\acrodef{AUC}{Area Under the Receiver Operator Characteristic}
\acrodef{DWT}{Discrete Wavelet Transform}
\acrodef{FFT}{Fast Fourier Transform}
\acrodef{TPOT}{Tree-based Pipeline Optimization Tool}
\acrodef{TUAR}{Temple University Artifact Corpus}
\acrodef{TUEV}{Temple University Event Corpus}
\acrodef{AEP}{Auditory Evoked Potential}

\acrodef{BSS}{Blind Source Separation}
\acrodef{ICA}{Independent Component Analysis}
\acrodef{ICs}{Independent Components}
\acrodef{ASR}{Artifact Subspace Reconstruction}
\acrodef{PCA}{Principal Component Analysis}
\acrodef{GAP}{Global Average Pooling}
\acrodef{FCN}{Fully Connected Networks}
\acrodef{MLP}{Multi-Layer Perceptron}
\acrodef{NAS}{Neural Architectural Search}
\acrodef{FP/h}{False Positives per Hour}
\acrodef{BVP}{Blood volume Pulse}
\acrodef{EDA}{Electrodermal Activity}
\acrodef{ACC}{Accelerometer}
\acrodef{CAE}{Convolutional Autoencoder}
\acrodef{SSWCE}{Sensitivity-Specificity Weighted Cross-Entropy}
\acrodef{CE}{Cross-Entropy}
\acrodef{PPG}{Photoplethysmography}
\acrodef{NIRS}{Near-infrared Spectroscopy}

\acrodef{SBP}{systolic Blood Pressure}
\acrodef{SVM}{Support Vector Machine}
\acrodef{LDA}{Linear Discriminant Analysis}
\acrodef{MSE}{Multiscale Entropy}
\acrodef{EMD}{Empirical Mode Decomposition}
\acrodef{HR}{Heart Rate}
\acrodef{GSR}{Galvanic Skin Response}
\acrodef{NFC}{Near Field Communication}
\acrodef{RR}{Respiration Rate}
\acrodef{RSP}{Respiration}
\acrodef{BCM}{Body Composition Measurement}
\acrodef{SKT}{Skin Temperature}
\acrodef{ICG}{Impedance Cardiography}
\acrodef{AFE}{Analog-Front-End}
\acrodef{SSI}{Silent speech interface}
\acrodef{IMU}{Inertial Measurement Unit}
\acrodef{EEG}{Electroencephalogram}
\acrodef{EOG}{Electrooculography}
\title{SilentWear: an Ultra-Low Power Wearable System for EMG-based Silent Speech Recognition}

\author{Giusy Spacone, \IEEEmembership{Graduate Student Member, IEEE}, Sebastian Frey, \IEEEmembership{Graduate Student Member, IEEE}, Giovanni Pollo, \IEEEmembership{Graduate Student Member, IEEE}, Alessio Burrello, \IEEEmembership{{Member, IEEE}}, Daniele Jahier Pagliari, \IEEEmembership{{Member, IEEE}}, Victor Kartsch, \IEEEmembership{Member, IEEE}, Andrea Cossettini \IEEEmembership{Senior Member, IEEE} and Luca Benini, \IEEEmembership{Fellow, IEEE}
\thanks{This work was submitted on 20 February 2026.}
\thanks{G. Spacone, S.Frey, V.Kartsch, A. Cossettini, and L. Benini are with the Integrated Systems Laboratory of ETH Z{\"u}rich, Z{\"u}rich, Switzerland (correspondence to: \texttt{gspacone@iis.ee.ethz.ch}).}
\thanks{L. Benini is also with the Department of Electrical, Electronic and Information Engineering (DEI), University of Bologna, Bologna, Italy.}
\thanks{G. Pollo, A.Burrello, D.J. Pagliari are with the Department of Control and Computer Engineering, Politecnico of Turin, Italy.}
}

\IEEEtitleabstractindextext{%
\fcolorbox{abstractbg}{abstractbg}{%
\begin{minipage}{\textwidth}%
\begin{wrapfigure}[12]{r}{3.6in}%
\includegraphics[width=3.5in]{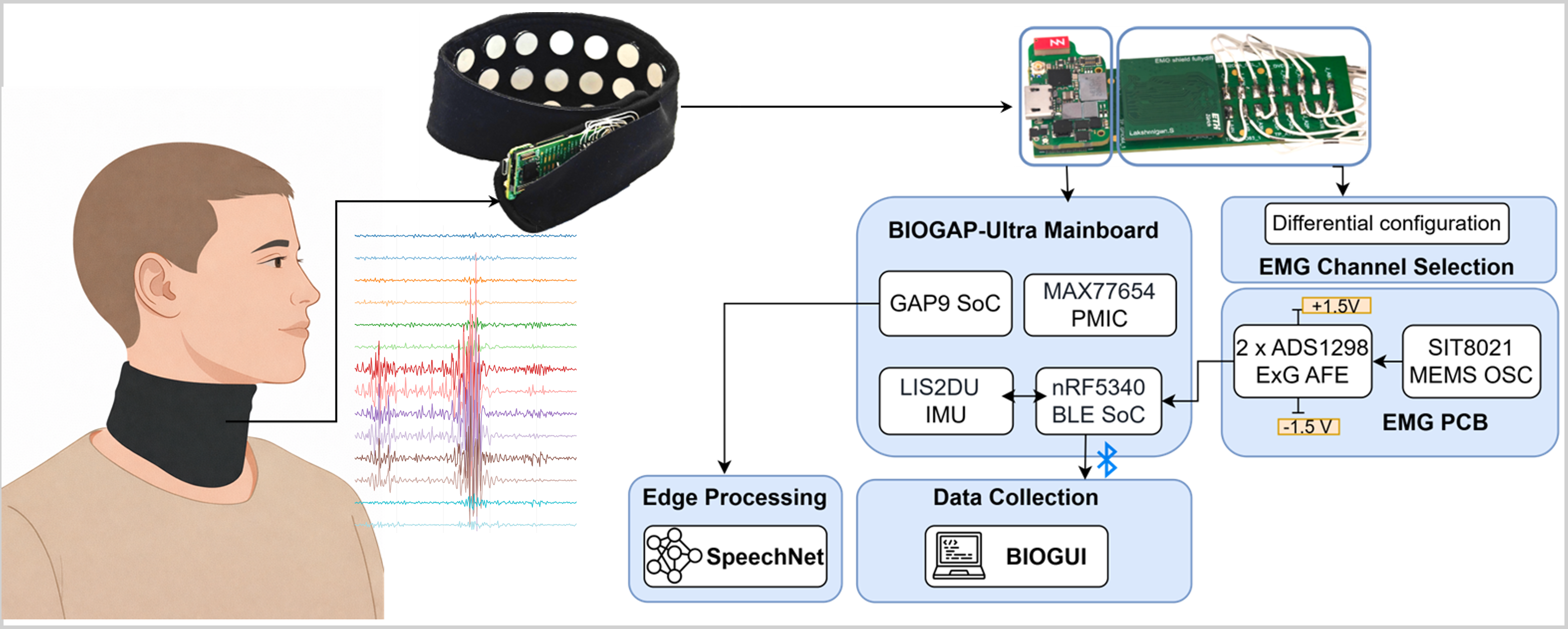}%
\end{wrapfigure}%
\begin{abstract}

Detecting speech from biosignals is gaining increasing attention due to the potential to develop \review{human-computer} interfaces that are noise-robust, privacy-preserving, and scalable for both clinical applications and daily use. However, most existing approaches remain limited by insufficient wearability and the lack of edge-processing capabilities, which are essential for minimally obtrusive, responsive, and private assistive technologies. 

In this work, we present \textit{SilentWear}, a fully wearable, textile-based neck interface for EMG signal acquisition and processing. Powered by BioGAP-Ultra, the system enables end-to-end data acquisition from 14 differential channels and on-device speech recognition. \textit{SilentWear} is coupled with \textit{SpeechNet}, a lightweight 15k-parameter CNN architecture specifically tailored for EMG-based speech decoding, achieving an average cross-validated accuracy of 84.8\textpm4.6\% and 77.5\textpm6.6\% for vocalized and silent speech, respectively, over eight representative human–machine interaction commands collected over multiple days.
We evaluate robustness to repositioning induced by the multi-day use. In an inter-session setting, the system achieves average accuracies of 71.1\textpm8.3\% and 59.3\textpm2.2\% for vocalized and silent speech, respectively. To mitigate performance degradation due to repositioning, we propose an incremental fine-tuning strategy, demonstrating more than 10\% accuracy recovery with less than 10 minutes of additional user data. Finally, we demonstrate end-to-end \review{real-time} on-device speech recognition on \review{a commercial multi-core microcontroller unit (MCU)}, achieving an energy consumption of 63.9~\textmu J per inference with a latency of 2.47~ms. With a total power consumption of 20.5~mW for acquisition, inference, and wireless transmission of results, SilentWear enables continuous operation for more than 27~hours.
\end{abstract}

\begin{IEEEkeywords}
edge-AI, EMG, HMI, on-device, online, speech, silent speech, ultra-low power, wearables
\end{IEEEkeywords}
\end{minipage}}}
\makeatletter
\def\ps@mynotice{%
  \def\@oddfoot{%
    \hfil
    \parbox[t]{\textwidth}{\centering\scriptsize
      \copyright\ This work has been submitted to the IEEE for possible
      publication. Copyright may be transferred without notice, after which
      this version may no longer be accessible.%
    }%
    \hfil
  }%
  \def\@evenfoot{\@oddfoot}%
}
\makeatother
\maketitle
\thispagestyle{mynotice} 
\pagestyle{mynotice}      
\vspace{-0.5cm}
\section{Introduction}\label{sec:intro}
\acp{SSI} are assistive technologies designed to decode intended speech without overt vocalization, aiming to restore communication for individuals with speech impairments and to enable silent interaction in everyday contexts, such as noisy environments or privacy-sensitive scenarios~\cite{tang_2026_sensing}. Their development is further driven by the increasing demand for efficient, robust, and socially acceptable assistive solutions, fostering innovation toward comfortable, miniaturized, and energy-efficient wearable systems~\cite{s22207784}. As these technologies mature and scale, they are also becoming more cost-effective, accelerating the transition of \acp{SSI} from research prototypes to practical real-world applications.

\acp{SSI} rely on various sensing modalities, among which surface \ac{EMG} is particularly attractive due to its ability to capture speech-related activity directly at the neuromuscular source, providing low-latency information~\cite{tang_2026_sensing, chowdhury_2025_decoding}. At the same time, \ac{EMG} is well suited for non-invasive, minimally obtrusive wearable systems with low energy requirements and able to operate over multiple days\review{\cite{kaifosh_2025_generic}}.

Advances in \ac{DL} models have become key enablers for practical \acp{SSI}, allowing the extraction of complex non-linear representations from \acp{EMG} signals and improving robustness to the non-stationarities encountered in daily use. However, realizing truly wearable \acp{SSI} requires shifting the computational burden from the cloud to the edge, demanding efficient on-device inference to ensure real-time operation and responsiveness while minimizing wireless communication and preserving data privacy~\cite{biomimetics10030166, tang_2026_sensing}. Recent developments in energy-efficient edge platforms, such as \review{VersaSens \cite{najafi_2024_versasens}} or BioGAP-Ultra~\cite{frey_2026_biogapultra}, address these constraints by providing efficient biosignal acquisition together with the computational capability required to execute \ac{DL} models under strict latency and battery limitations.

Despite steady progress and \review{partial} advancements, current EMG-based \acp{SSI} still fall short in several key areas essential to reliable, socially acceptable everyday use~\cite{9205294}. The first bottleneck is the sensing interface: most of existing systems still rely on wet electrodes, facial placements, or bulky form factors, which are cumbersome, stigmatizing, and impractical for continuous wear. 
Second, while some approaches report strong results on individual subjects, systematic characterization across diverse users and, critically, performance stability across sessions and multiple days remain limited. 
Third, introducing \acp{SSI} into daily life requires embedded deployment, which guarantees low latency, \review{long} battery life and inherently safeguards user privacy by keeping sensitive biosignals local \cite{tang_2026_sensing}.

To cope with these challenges, we introduce \emph{SilentWear}, a novel, fully dry, textile-based speech interface in a neckband form factor, powered by the BioGAP-Ultra platform~\cite{frey_2026_biogapultra} for end-to-end on-device \ac{EMG} acquisition and processing. Extending our previous conference work~\cite{meier_2025_parallel}, we present the following additonal novel contributions:

\begin{itemize}
    \item We present a new dataset\footnote{\label{foot:dataset}{Dataset: {\scriptsize\url{https://huggingface.co/datasets/PulpBio/SilentWear}} \\ Code: {\scriptsize\url{https://github.com/pulp-bio/SilentWear}}}}from four subjects, consisting of vocalized and silent EMG recordings of 8 commands for \acp{HMI}, collected over multiple days, under realistic sensor repositioning scenarios. 
    \item We propose \textit{SpeechNet}, a compact \ac{CNN} architecture, showcasing the feasibility of detecting \ac{HMI}-related commands on-board, with an average cross-validated accuracy of $84.8\pm4.6\%$ for vocalized and $77.5\pm6.6\%$ for silent speech.
    \item We assess the cross-day generalization by evaluating on an unseen recording day (excluded from training), achieving an average inter-session accuracy across subjects of $71.1\pm8.3\%$ and $59.3\pm2.2\%$ (vocalized/non-vocalized).
    \item To overcome the challenges arising from repositioning, we propose an incremental fine-tuning strategy, achieving more than 10\% accuracy recovery with less than $10\text{minutes}$ of data re-collection.
    \item We deploy the network architecture on the GAP9 \ac{MCU} embedded in BioGAP-Ultra, showcasing an inference time of $2.47\text{ms}$, with an energy/inference of $63.9$ $\mu\text{J}$. Including signal acquisition, on-device inference, and wireless transmission of classification results, the system consumes $20.5\,\text{mW}$, yielding $27.1\,\text{h}$ of battery life with a $150\,\text{mAh}$ \review{Li-Po} battery.
\end{itemize}

\section{Related Works}\label{sec:related}
Over the last decade, significant progress has been made towards the development of wearable \acp{SSI}, leveraging diverse sensing modalities such as \ac{IMU}, strain sensing, \ac{EEG}, and \ac{EOG}. Among these, EMG is particularly promising due to its ability to capture neuromuscular activations at the source while remaining cost-effective, easy to use, and minimally obtrusive~\cite{tang_2026_sensing}.

On the algorithmic side, open-source datasets such as that of Gaddy \textit{et al.}~\cite{gaddy_2020_digital} have enabled the development of robust deep learning architectures tailored to speech recognition. However, such datasets have been collected under controlled laboratory conditions that do not reflect realistic daily usage (e.g., obtrusive electrode placements, wet electrodes, fixed sensor positioning).

Bridging the gap between algorithmic demonstrations under ideal data collection settings and truly wearable systems requires addressing \review{major} challenges:
i) the development of unobtrusive sensing interfaces robust to daily variability induced by device removal, repositioning, and physiological changes;
ii) the adoption of energy-efficient hardware platforms supporting on-device processing for low latency, extended battery life, and privacy preservation;
iii) the design of lightweight algorithms compatible with embedded deployment constraints.

Several wearable \acp{SSI} solutions have been proposed in the \ac{SoA} and are discussed in the following sections; a comparative overview is provided in Table~\ref{tab:ssi_comparison_transposed}.

Kapur \textit{et al.}~\cite{kapur_alterego_2018} presented one of the first approaches toward wearable SSIs with AlterEgo, a head-worn system employing face-contact electrodes. They demonstrated strong recognition performance (92\% accuracy) on a dataset collected from 10 subjects. Despite these promising results, the system relies on a relatively bulky form factor, which may limit user acceptance in everyday scenarios. To address such \review{limits}, subsequent works, including Liu \textit{et al.}~\cite{liu_2020_epidermal} and Wang \textit{et al.}~\cite{wang_2021_all}, focused on the development of novel electrode technologies (e.g., tattoo-like electrodes) designed for conformal facial and jaw placement. These approaches reduced thickness and improved flexibility and comfort. Nevertheless, face-based electrode placement \review{raises stigmatization concerns} and may hinder long-term usability and social acceptance. 

More recently, alternative sensing locations have been explored to improve wearability. Wu \textit{et al.}~\cite{wu_towards_2024} proposed a textile-based neckband integrating gold-plated electrodes and an energy-efficient acquisition system (WANDmini~\cite{zhou_2019_wandmini}) with \ac{BLE} connectivity for multi-day operation ($46\,\mathrm{mW}$). They demonstrated the feasibility of recognizing 11 phonemes from two subjects, achieving 92.7\% accuracy with a Random Forest classifier. However, the results are based on an extremely limited dataset (110 samples in total), a vocabulary with limited real-world applicability, and the work remains narrow in its algorithmic exploration. Tang \textit{et al.}~\cite{tang_2025_wireless} investigated the ear region, embedding four textile-based electrodes into commercial headphone earmuffs. The system, which relies on custom hardware (a SeedStudio EMG readout combined with a WiFi-enabled ESP32-S3 MCU), achieves 96\% recognition accuracy on 10 daily-life commands (e.g., start, stop) using a Squeeze-Excitation ResNet over a dataset comprising 10 subjects.

Despite significant progress, several challenges remain before EMG-based wearable SSIs can achieve widespread real-world adoption. First, most algorithmic studies conducted on data collected with wearable \acp{SSI} rely on small datasets collected under highly controlled conditions, typically excluding realistic factors such as sensor repositioning and multi-day acquisitions. These factors, however, are intrinsic to daily use and critically affect system robustness. Second, the lack of standardized data collection protocols and evaluation methodologies makes it difficult to conduct fair algorithmic comparisons across studies. Third, the majority of evaluations are performed on external computing platforms (laptops). For truly deployable systems, low-latency and privacy-preserving operation, on-device inference under embedded hardware constraints is needed.

Our work targets the identified gap at the intersection of wearability, robustness, and embedded intelligence, extending the results presented in our previous work. We explicitly address the challenges arising from multi-day system use and enable fully end-to-end, low-latency, and privacy-preserving on-device speech detection.

\section{Materials and Methods}
\label{sec:methods}
\subsection{Silent Wear Interface}
\begin{figure}
    \centering
    \includegraphics[width=1\linewidth]{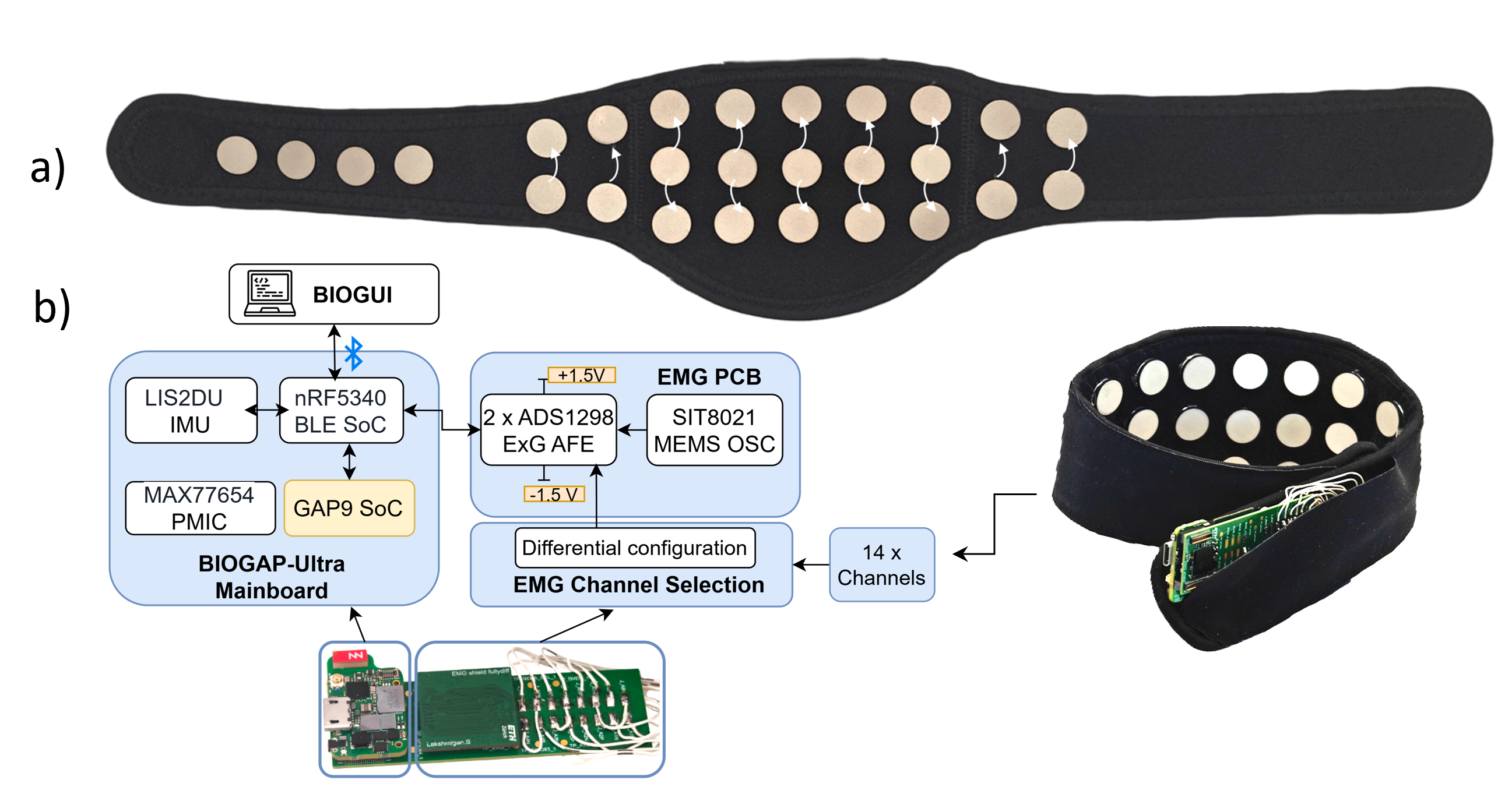}
    \caption{a) SilentWear neckband. The interface features 14 \ac{EMG} differential acquisition channels; 4 channels provide the ground reference.; b) Overview of the acquisition hardware, based on BioGAP-Ultra. The system features a baseboard, for \ac{BLE} communication and on-device processing, and an EMG acquisition board, with a total size of $26\times65\times13\,mm^{3}$.}
    \label{fig:hw}
    \vspace{-0.6cm}
\end{figure}
In this study, we employ the hardware presented in our previous work \cite{meier_2025_parallel}. A description of the system design is presented in \mbox{Fig. \ref{fig:hw}}.

The interface is a soft-fabric neckband with fully dry electrodes, adjustable in size via velcro straps. The design follows the principles of simplicity and everyday usability. Since no wet electrodes or conductive gel are required, it can be worn immediately, supporting daily use without being intrusive. The inner side of the band contains 27 sewn-in snap fasteners that support connection to 27 fully dry electrodes (Datwyler, SoftPulse \cite{datwyler_softpulse}). The central portion of the neckband contains 15 electrodes arranged in three rows and connected as overlapping differential channels (top-middle and middle-bottom), such that adjacent channels share the middle electrode, resulting in 10 effective differential channels. This design choice ensures higher coverage of the infrahyoid muscles, which are primary active during phonation and articulation \cite{lang_anatomy_esoph}. In addition, four fully differential channels are placed laterally (two per side of the neck), to ensure coverage of the sternocleidomastoid muscle \cite{puig_2024_characterization}. Hence, the system features a total of 14 acquisition channels; four electrically shorted electrodes, placed at the back of the neck, provide the ground reference. 

The acquisition and processing hardware is based on BioGAP-Ultra \cite{frey_2026_biogapultra}.
BioGAP-Ultra comprises a baseboard that integrates GAP9, a \ac{SoA} \ac{ULP} \ac{SoC} with AI capabilities, able to execute \acp{DNN} efficiently on-board and the Nordic nRF5340 \ac{MCU} with \ac{BLE} capabilities. 
The baseboard interfaces with a biopotential acquisition board equipped with two Texas Instruments ADS1298 \acp{AFE}, which enable simultaneous acquisition from 16 differential EMG channels. A complete description of the system is provided in \cite{frey_2026_biogapultra}. The acquisition hardware has a total dimension of $26\times65\times13\,mm^{3}$ and is powered by a standard $150\,\text{mAh}$ Li-Po battery. 
\vspace{-0.3cm}
\subsection{Data Collection Protocol}\label{subsec:datacollection}
\begin{figure}[H]
    \centering
    \includegraphics[width=1\linewidth]{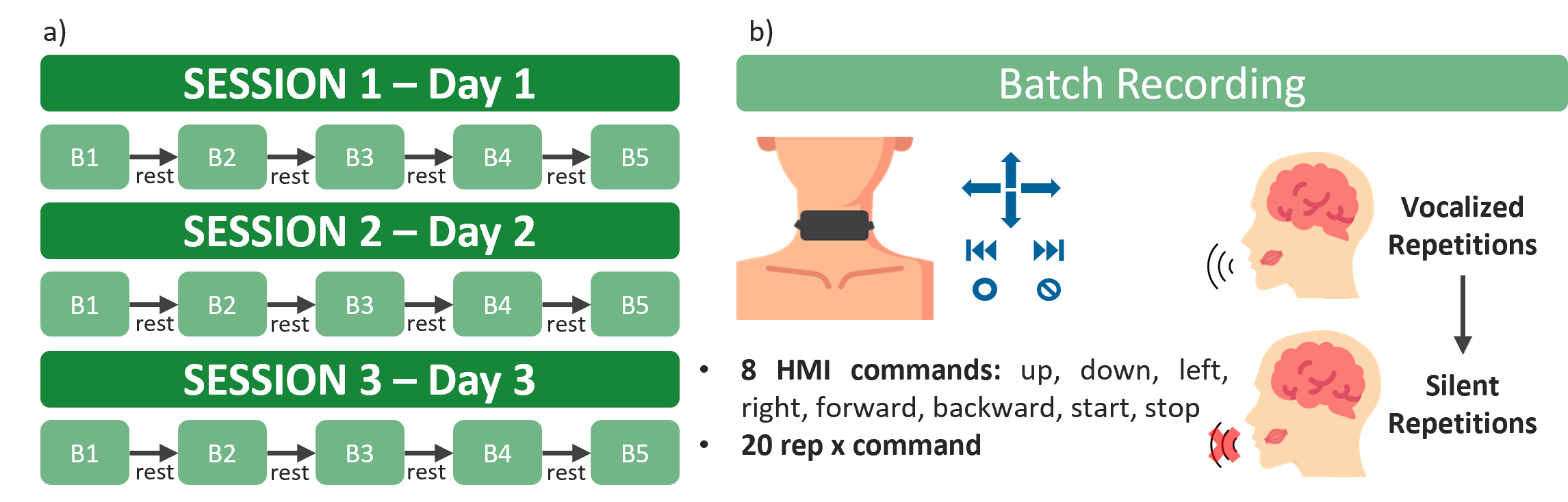}
    \caption{Data collection protocol. 
    (a) Recordings are organized into sessions conducted over multiple days. Each session consists of 10 batches (5 vocalized and 5 silent), executed alternately, with 2-minute rest intervals between batches. 
    (b) In each batch, subjects repeat 8 HMI-related commands, each performed 20 times in randomized order.}
    \label{fig:protocol}
    \vspace{-0.2cm}
\end{figure}

We adopt a data collection protocol inspired by our previous work~\cite{meier_2025_parallel}. An overview of the protocol is shown in Fig.~\ref{fig:protocol}.

Data acquisition and user guidance are managed through the open-source BioGUI software~\cite{orlandi_biogui}\footnote{\footnotesize{Data collection software: {\scriptsize\url{https://github.com/pulp-bio/biogui/tree/sensors_speech}}}}. Specifically, the BioGUI is responsible for setting the acquisition configuration (sampling rate of $500~\text{Hz}$ and \ac{PGA} gain of 6) for the two ADS1298 \acp{AFE}, displaying real-time EMG signals, displaying user instructions, labeling and storing data.

During data collection, participants sit comfortably in a chair. The recordings are organized into 3 \textit{sessions} per subject, conducted over multiple days. Between sessions, the neckband is repositioned to reproduce realistic usage conditions and to account for variability in sensor placement. Each session consists of multiple \textit{batches}. In each batch, subjects are instructed to repeat eight HMI-related commands (\textit{up, down, left, right, forward, backward, start, stop}) in randomized order. Each command is repeated $20$ times. For every repetition, subjects are given $2\,\text{s}$ for speech production (vocalized or silent), followed by a $1.5\,\text{s}$ rest interval. During the rest period, the GUI displays the next command to minimize reaction time. Each session comprises ten batches, alternating between five vocalized and five silent batches. A $2\,\text{min}$ break is provided between consecutive batches to reduce fatigue.

With $8$ commands, $20$ repetitions per command, and $3.5\,\text{s}$ per repetition ($2\,\text{s}$ production $+$ $1.5\,\text{s}$ rest), each batch lasts approximately $560\,\text{s}$ ($\approx 9\,\text{min}\,20\,\text{s}$). Including the $2\,\text{min}$ breaks between the ten batches, the total duration of each session is approximately $110\,\text{min}$. Overall, the dataset comprises 2400 vocalized and 2400 silent utterances per subject ($8\,\text{commands} \times 20\,\text{repetitions} \times 5\,\text{batches} \times 3\,\text{sessions}$). For each condition, an additional 2400 samples correspond to rest segments.

We collected data from four subjects (three male, one female, average age 30 years old). Written informed consent was obtained from all participants prior to data acquisition. All procedures complied with the Declaration of Helsinki.
\vspace{-0.3cm}
\subsection{Data Processing and Dataset Creation}
Raw EMG signals are preprocessed using a fourth-order zero-phase Butterworth high-pass filter with a cutoff frequency of $20\,\mathrm{Hz}$, followed by a $50\,\mathrm{Hz}$ notch filter, consistent with our previous work.

Labels are derived from the trigger signal generated by the BioGUI and recorded together with the EMG data. Each word and each rest period is assigned a unique identifier.
The continuous EMG signal is segmented into windows starting at the corresponding trigger onset and ending at the subsequent trigger (rest between words). Each window inherits the label of the associated word or rest period.
As a result, for each subject and condition (silent or vocalized), the dataset contains 4800 samples: 300 samples per command (8 commands) and 2400 rest samples. During model training, the rest class is randomly downsampled to ensure class balance.
\vspace{-0.2cm}
\subsection{Evaluation Settings}\label{subsec:eval_methods}
To assess the quality of \review{the acquired signals}, we tested three evaluation protocols to discriminate between the 9 classes (8 commands plus rest) collected:

\textit{I) Global Evaluation Setting}: Data from all sessions of a subject are pooled and evaluated using a \textit{leave-one-batch-out} cross-validation scheme. In each fold, four batches from each session are used for training, and the remaining three batches (one for each session) are used for testing.
This protocol establishes a baseline and serves as a means of comparison with prior works, which do not include sensor repositioning. Since the dataset spans multiple recording days, both training and test sets draw from the same overall data distribution. Testing is, however, always performed on unseen batches, thereby explicitly accounting for batch-to-batch variability.

\textit{II) Inter-Session Evaluation Setting}: This setting adopts a \textit{leave-one-session-out} cross-validation scheme. In each fold, two sessions (total of 10 batches) are used for training, and the remaining session for that subject is used for testing. Therefore, the held-out test fold contains completely unseen data, collected on a different day and after sensor repositioning. 
This protocol assesses model robustness to sensor repositioning and day-to-day variability. 

\textit{III) Incremental Fine-Tuning Setting}: This protocol assesses the amount of data needed to calibrate the system after sensor repositioning by an incremental training strategy, which is studied over two scenarios:
    \begin{itemize}
    \item \textit{a) Fine-tuning from a pre-trained model}. 
    We initialize the model with weights pre-trained on two sessions (as in Setting \textit{II}) and evaluate its zero-shot accuracy on the first batch of a new acquisition session. The model is then incrementally fine-tuned by incorporating data from the new batches. Specifically, 70\% of the batch data (126 utterances, corresponding to 14 samples per class) are used for fine-tuning, while the remaining 30\% are reserved for validation. The updated model is then evaluated zero-shot on the subsequent batch, followed by another fine-tuning step. This procedure is repeated sequentially for the four batches.
    We repeat this experiment three times, using 2 sessions each time for model pre-training (as in \textit{II}) and fine-tuning on the remaining one.

    \item \textit{b) Training from scratch}. 
    We initialize the model's weights randomly. At each training round, the model is trained using 70\% of a new batch (126 utterances, 14 samples per class), while the remaining 30\% are used for validation. Performance is evaluated in a zero-shot manner on the next unseen batch (from batch 2 to batch 5).
    We repeat this experiment for the three acquisition sessions.
\end{itemize}

We train subject-specific and condition-specific (vocalized or silent) models.
For all evaluation settings, we report the mean and standard deviation of the balanced accuracy across batches or sessions. In addition, we report results averaged across subjects by computing the mean of the per-subject scores and their corresponding standard deviation.
\vspace{-0.3cm}
\subsection{Network Architecture: SpeechNet}
We design a network architecture inspired by EpiDeNet \cite{ingolfsson_2023_epidenet}, developed for seizure detection and already proven to be suitable for \ac{HMI}-related tasks \cite{frey_2024_gapses}. 

As in \cite{ingolfsson_2023_epidenet}, the first three blocks learn temporal patterns using kernels that operate independently on each sensor channel. For the first three layers, zero-padding is applied to make the architecture flexible with respect to any temporal window size.
The last two blocks learn cross-channel spatial representations, integrating information across multiple channels. The features learned in the convolutional blocks are then pooled and passed to a single fully connected layer (output size 9 classes). Further details on the network architecture are displayed in Table \ref{tab:network_architecture}. The total number of parameters is 15489. 

\begin{table}[t]
\centering
\caption{SpeechNet Architecture}
\label{tab:network_architecture}
\begin{tabular}{c c c c c}
\hline
\textbf{Type} & \textbf{Layer} & \textbf{\# Filters} & \textbf{Kernel} & \textbf{Output} \\
\hline
$\phi^{1}$ & Conv2D & 8  & $(1 \times 4)$  & $(8, 14, T)$ \\
           & MaxPool & -- & $(1 \times 8)$ & $(8, 14, T/8)$ \\
\hline
$\phi^{2}$ & Conv2D & 16 & $(1 \times 16)$ & $(16, 14, T/8)$ \\
           & MaxPool & -- & $(1 \times 4)$  & $(16, 14, T/32)$ \\
\hline
$\phi^{3}$ & Conv2D & 16 & $(1 \times 8)$  & $(16, 14, T/32)$ \\
           & MaxPool & -- & $(1 \times 4)$  & $(16, 14, T/128)$ \\
\hline
$\phi^{4}$ & Conv2D & 32 & $(7 \times 1)$  & $(32, 8, T/128)$ \\
           & MaxPool & -- & $(1 \times 1)$  & $(32, 8, T/128)$ \\
\hline
$\phi^{5}$ & Conv2D & 32 & $(7 \times 1)$  & $(32, 2, T/128)$ \\
           & MaxPool & -- & $(1 \times 1)$  & $(32, 2, T/128)$ \\
\hline
$\phi^{6}$ & AdaptiveAvgPool & -- & -- & $(32, 1, 1)$ \\
\hline
$\phi^{7}$ & Dense & -- & -- & $9$ \\
\hline
\end{tabular}
\begin{tablenotes}
  \small
  \item T = number of time samples; for $\phi^{1}, \phi^{2}, \phi^{3}$ zero-padding is applied on the time dimension.
\end{tablenotes}
\vspace{-0.5cm}
\end{table}

The network is trained using the Cross-Entropy Loss and the Adam optimizer. 
For all evaluation settings, we use an initial learning rate of $10^{-3}$ and the \textit{Reduction on Plateau} scheduler, applied to the validation loss (patience of 2 epochs). Weight decay is set to $10^{-4}$. 
In evaluation settings I, II, III-b, we train for a maximum of 100 epochs. For evaluation setting III-a, we apply fine-tuning for 50 epochs on models pre-trained on 100 epochs (as in II). In both cases, early stopping with a patience of 10 epochs is applied if validation loss stops improving. Models are initially trained with an EMG input size of $1400\text{ms}$, as in our previous work \cite{meier_2025_parallel}, and to enure that the same amount of samples is available both for words and the rest data (duration of $1500\text{ms}$ between words, as described in Sec. \ref{subsec:datacollection}).

In addition, to enable comparison with \cite{meier_2025_parallel} and assess the performance of the new architecture, we also train a Random Forest Classifier (maximum 100 estimators), adopting manual feature extraction. The $1400\text{ms}$ of data are segmented into seven non-overlapping windows; from each window, we extract time and frequency-domain features and use them as training data. Additional details are available in \cite{meier_2025_parallel} (Section II-D). 

We use PyTorch for the \ac{CNN} models; the Scikit-learn library is used to train the Random Forest models and for data preparation.
We run all experiments on a single GPU (Nvidia GEFORCE GTX 1080).
\vspace{-0.3cm}
\subsection{Ablation on window size and information transfer rate}
We conduct an ablation on the EMG window size to assess its impact on two scores: accuracy and \ac{ITR}, a widely adopted metric in brain-computer interfaces \cite{itr_definition}.
The ITR (bit/min) is computed as: 
\begin{equation*}
    \mathrm{ITR} = \frac{60}{T}
    \left[
    \log_2(C)
    + P \log_2(P)
    + (1 - P)\log_2\!\left(\frac{1 - P}{C - 1}\right)
    \right]
\end{equation*}
where $C$ is the number of classes, $T$ the window lenght (s), $P$ the classification accuracy \review{(0-1 range).}
We ablate over window sizes spanning from $400\text{ms}$ to $1400\text{ms}$, with $200\text{ms}$ steps.
\vspace{-0.3cm}
\subsection{Model Deployment}
We deploy the model on the GAP9 \ac{MCU} integrated in BioGAP-Ultra. GAP9 combines a single RISC-V controller core with a programmable 9-core RISC-V compute cluster and a dedicated Neural Network (NN) hardware accelerator (NE16). GAP9 provides a hierarchical on-chip memory system with $128\,\text{kB}$ of L1 memory and $1.6\,\text{MB}$ of L2 RAM, while BioGAP-Ultra augments the platform with an additional $64\,\text{MB}$ off-chip PSRAM used as L3 memory. 

For model deployment, we use the proprietary NNTool framework~\cite{nntool}. First, we export the trained network to the Open Neural Network Exchange (ONNX) format and import it into NNTool to map the computational graph to the GAP9 processor and generate optimized C code. NNTool also performs 8-bit integer post-training quantization for efficient execution.
Model constants (weights, biases, and quantization parameters) are stored in off-chip Flash (overall footprint $\approx$\SI{15.13}{kB}) and promoted to L2 at initialization. Since the complete network fits in on-chip memory, inference proceeds using only on-chip L2 storage, with per-layer tiles moved to L1 by the DMA during kernel execution.
We evaluate two operating points that reflect common deployment objectives:
\begin{itemize}
    \item \textit{Maximum inference speed}: NE16 runs at its highest frequency \SI{370}{MHz} and \SI{0.8}{V} to minimize latency.
    \item \textit{Highest energy efficiency:} NE16 runs at \SI{240}{MHz} and \SI{0.65}{V}, its most efficient operating point.
\end{itemize}

\section{Results}\label{sec:res}
\vspace{-0.3cm}
\subsection{EMG acquisitions}
\begin{figure*}
    \centering
    \includegraphics[width=1.0\linewidth]{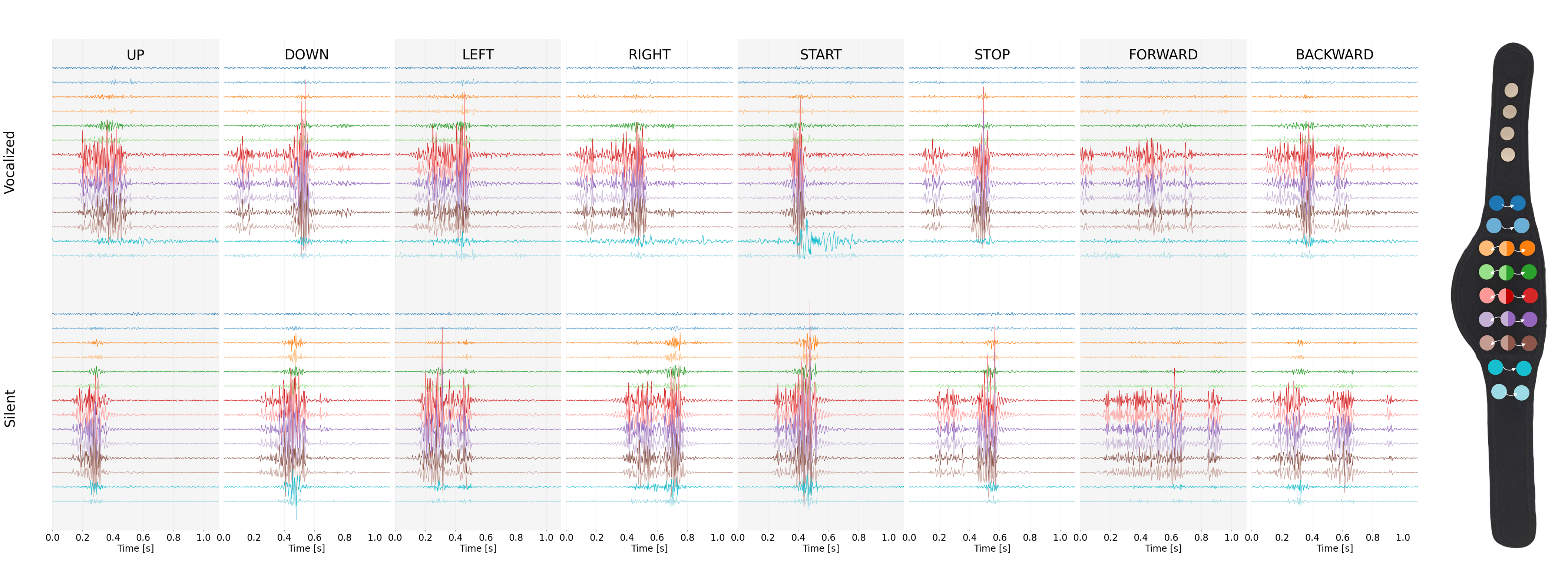}
    \caption{Visualisation of the signals of the 14 EMG Channels for the 8 HMI commands, for vocalized (top) and silent (bottom) recordings.}
    \label{fig:emg_signals}
\end{figure*}

Figure~\ref{fig:emg_signals} shows a representative example of the signals acquired from one of the subjects (S04) during the repetition of each command included in the dataset, for both vocalized and silent speech.
Clear differences \review{can be visually observed} between the rest (no speech) and articulation phases in both vocalized and silent conditions, confirming the system's ability to detect speech-related muscle activity reliably.

Furthermore, across all commands, the central channels exhibit a higher signal-to-noise ratio (SNR) than the lateral channels. Based on the neckband design, these central electrodes are positioned over the infrahyoid muscles (sternohyoid, omohyoid, sternothyroid, and thyrohyoid), which are involved in laryngeal stabilization and vertical movement during phonation and articulation~\cite{lang_anatomy_esoph}. The stronger activation observed on these channels is therefore consistent with the physiological role of these muscles in modulating vocal fold positioning and supporting speech production.

\vspace{-0.3cm}
\subsection{SpeechNet Performance}
\begin{table*}[t]
\centering
\caption{Accuracy Comparison for Global and Inter-Session Evaluations between Random Forest and SpeechNet (window size 1.4s).}
\label{tab:results_merged_shared}
\scriptsize
\setlength{\tabcolsep}{2.5pt}

\begin{tabular}{ll|ccccc|ccccc}
\toprule

& & \multicolumn{5}{c}{\textbf{Global Evaluations}} & \multicolumn{5}{c}{\textbf{Inter-Session Evaluations}} \\
\cmidrule(lr){3-7} \cmidrule(lr){8-12}
Condition & Model 
& S01 & S02 & S03 & S04 & Avg. 
& S01 & S02 & S03 & S04 & Avg. \\

\midrule
\multirow{2}{*}{Vocalized} & Random Forest
& $78.4 \pm 3.0$ & $60.1 \pm 3.0$ & $63.2 \pm 4.8$ & $75.9 \pm 4.7$ & $69.4 \pm 7.9$
& $70.7 \pm 2.6$ & $42.4 \pm 14.1$ & $53.1 \pm 8.6$ & $67.9 \pm 7.0$ & $58.5 \pm 11.5$ \\

& SpeechNet
& $90.3 \pm 1.6$ & $82.6 \pm 1.3$ & $78.3 \pm 6.2$ & $87.8 \pm 2.5$ & $84.8\pm 4.6$
& $77.0 \pm 6.2$ & $58.8 \pm 19.0$ & $68.3 \pm 2.5$ & $80.2 \pm 7.4$ & $71.1 \pm 8.3$ \\

\midrule
\multirow{2}{*}{Silent} & Random Forest
& $68.2 \pm 3.1$ & $64.8 \pm 3.3$ & $59.6 \pm 3.0$ & $58.7 \pm 2.7$ & $62.9 \pm 3.9$
& $56.2 \pm 3.2$ & $47.4 \pm 8.6$ & $45.6 \pm 4.5$ & $45.6 \pm 1.8$ & $48.7 \pm 4.4$ \\
& SpeechNet
& $84.4 \pm 3.2$ & $83.2 \pm 2.0$ & $74.1 \pm 2.9$ & $68.5 \pm 2.8$ & $77.5 \pm 6.6$
& $62.9 \pm 2.3$ & $56.8 \pm 17.7$ & $59.2 \pm 2.8$ & $58.5 \pm 3.2$ & $59.3 \pm 2.2$ \\
\bottomrule
\end{tabular}
\vspace{-0.5cm}
\end{table*}

To \review{analyze} the performance gains of SpeechNet over our previous work, we first compare it against the previously used Random Forest classifier. To ensure a direct comparison, both models are trained using a window length of $1400\,\mathrm{ms}$. Table~\ref{tab:results_merged_shared} summarizes the performance for the two pipelines under the \textit{Global} and \textit{Inter-Session} evaluation settings, where SpeechNet consistently outperforms the Random Forest baseline in both settings and speaking conditions. Specifically, in the \textit{Global} setting, SpeechNet improves accuracy by approximately 15\% for vocalized speech and 15\% for silent speech. In the \textit{Inter-Session} setting, the corresponding gains are approximately 13\% for vocalized and 11\% for silent speech. Given its consistently superior performance, SpeechNet is adopted as the reference model for subsequent analyses.

\begin{figure*}
    \centering
    \includegraphics[width=\textwidth]{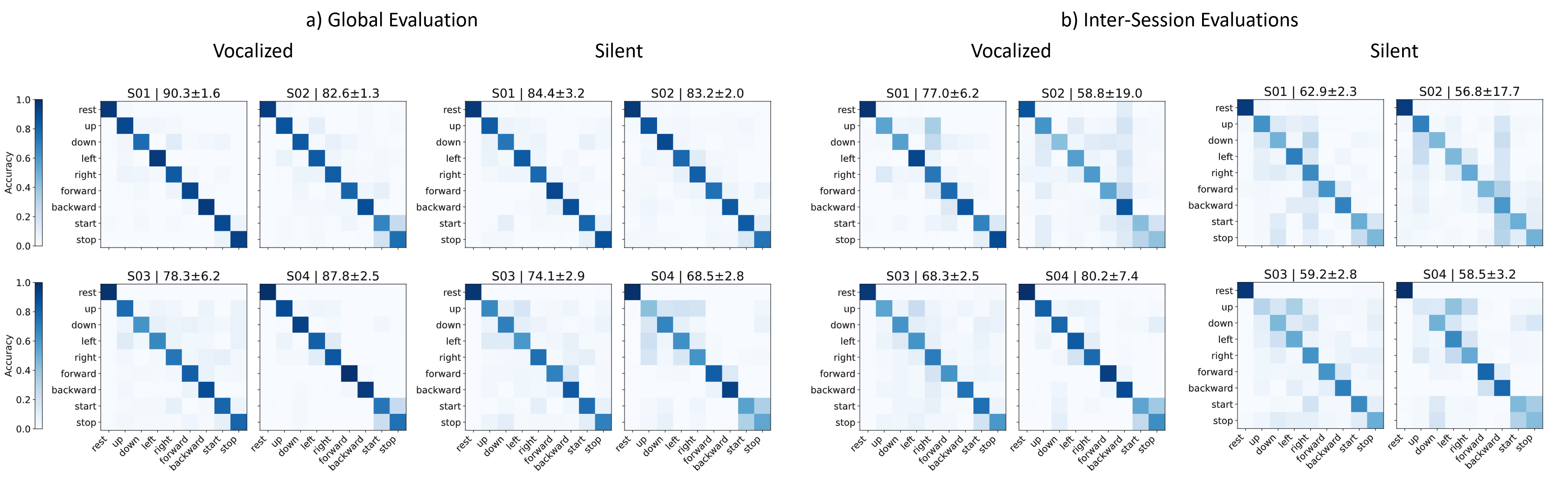}
    \caption{Per Subjects Confusion Matrices for the a) Global Evaluation Setting and b) Inter-Session Evaluation Setting for vocalized and silent speech, for EMG window size of $1400\text{ms}$.}
    \vspace{-0.5cm}
    \label{fig:cm_all}
    
\end{figure*}

Figure~\ref{fig:cm_all} reports the per-subject confusion matrices for the $1400\,\mathrm{ms}$ window under both the \textit{Global} and \textit{Inter-Session} evaluation settings. Several observations emerge.
In the \textit{Global} setting, diagonal dominance is consistently strong across subjects. In the vocalized condition, misclassifications are mainly confined to the \textit{start} and \textit{stop} pair. In the silent condition, errors predominantly affect short-duration commands, particularly for subjects S03 and S04, suggesting increased sensitivity to reduced articulatory dynamics.

Greater attention must be paid to the \textit{Inter-Session} evaluation case. First, the \textit{rest} class is classified with near-perfect accuracy across all subjects and speaking conditions, confirming the robustness of speech activity detection. This remains true even in the presence of short-duration commands (e.g., \textit{up}), which could potentially increase the risk of misclassification. Second, longer commands (e.g., \textit{forward}, \textit{backward}) exhibit, on average, higher recognition accuracy than shorter commands. When misclassifications occur, \textit{forward} and \textit{backward} are more likely to be confused with each other, plausibly due to similarities in their underlying neuromuscular activation patterns reflected in comparable EMG temporal dynamics and signal duration (see Fig.~\ref{fig:emg_signals}). 

For the remaining commands, no consistent cross-subject misclassification trend is observed. Off-diagonal entries display relatively high variance across folds, indicating that errors are not systematic but rather session-dependent. Such variability is likely driven by inter-session factors, including sensor repositioning, day-to-day physiological changes, and natural variations in speech tempo and articulation strategy.

In both the vocalized and silent conditions, subject S02 exhibits consistently lower performance than the other participants under the \textit{Inter-Session} setting, particularly in the vocalized condition. The reduced accuracy is primarily driven by the model trained on sessions 2-3 and evaluated on session 1. Visual inspection revealed differences in amplitude and waveform morphology in session 1 compared to subsequent sessions, which are plausibly attributable to differences in neckband fit (e.g., looser contact) or sensor placement. Importantly, this behavior reflects realistic usage conditions and highlights the sensitivity of inter-session performance to hardware fitting and positioning.

Overall, the performance gap between the \textit{Global} and \textit{Inter-Session} settings provides strong evidence that multi-day use induces a distribution shift in the EMG feature space. When training and testing samples are drawn from the same global distribution, the model can internalize session variability; conversely, when evaluated on a previously unseen session, performance degrades, underscoring the impact of sensor repositioning and day-to-day physiological variability, and highlighting the need for mitigation strategies, which will be further described in Section \ref{subsec:ft_results}.

\begin{figure}
    \centering
    \includegraphics[width=0.5\textwidth]{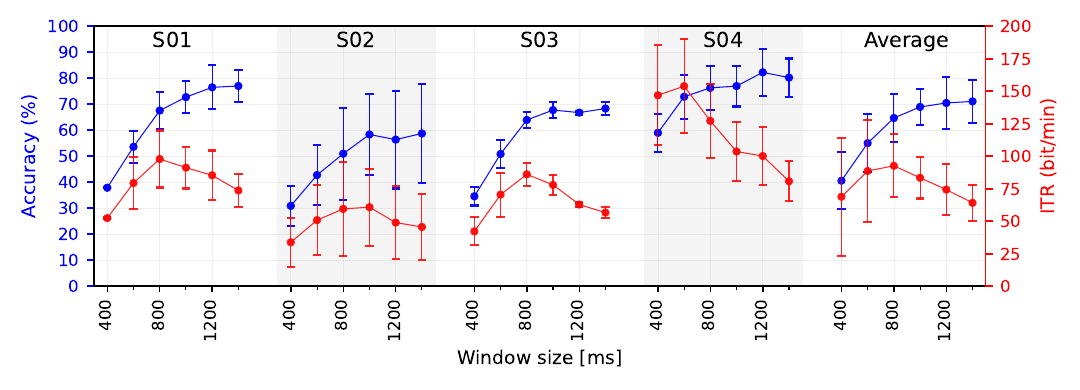}
    \vspace{-0.8cm}
    \caption{\ac{ITR} analysis for vocalized speech under the \textit{Inter-Session} evaluation setting. Blue curves (left axis) show mean accuracy and red curves (right axis) show mean ITR as a function of window size. For S01–S04, markers with error bars represent the mean and standard deviation across sessions. In the average plot, markers with error bars represent the mean and standard deviation across subjects.}
    \label{fig:itr_vocalized}
\end{figure}
\begin{figure}
    \centering
    \includegraphics[width=0.5\textwidth]{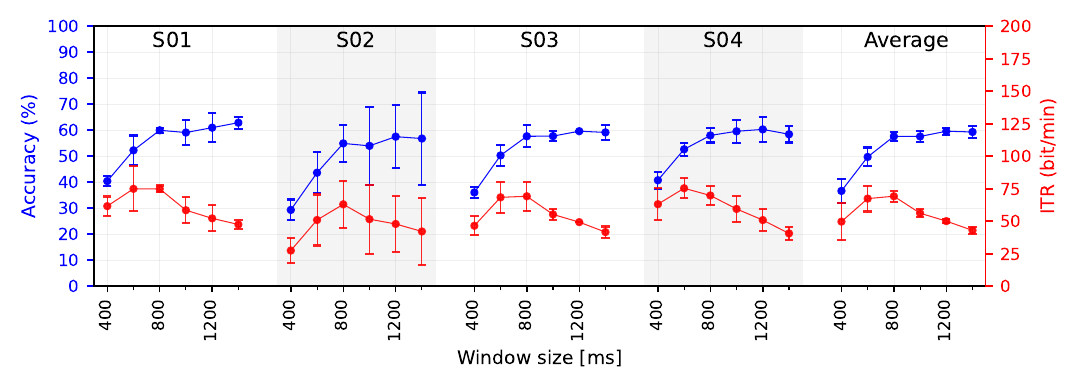}
    \vspace{-0.8cm}
    \caption{\ac{ITR} analysis for silent speech under the \textit{Inter-Session} evaluation setting. Plot structure and statistical reporting follow the same conventions as in Fig.~\ref{fig:itr_vocalized}.}
    \label{fig:itr_silent}
    \vspace{-0.5cm}
\end{figure}
\vspace{-0.3cm}
\subsection{ITR Analysis}
Figures~\ref{fig:itr_vocalized} and~\ref{fig:itr_silent} report the accuracy and \ac{ITR} under the \textit{Inter-Session} evaluation protocol for vocalized and silent speech, respectively.
For both conditions, larger window sizes yield the best results. 

In the vocalized condition, the highest average accuracy is obtained with a $1400\,\mathrm{ms}$ window (71.1\%). The maximum ITR is achieved with an $800\,\mathrm{ms}$ window, reaching $92.7\,\mathrm{bit/min}$.
In the silent condition, the highest average accuracy is obtained with a $1200\,\mathrm{ms}$ window (59.6\%). The maximum ITR is achieved with an $800\,\mathrm{ms}$ window, with $69.2\,\mathrm{bit/min}$.

In both speaking conditions, larger window lengths generally lead to higher classification accuracy. However, a shorter window ($800\,\mathrm{ms}$) provides a practical trade-off between robustness (near-peak accuracy), throughput (high ITR), and reduced latency, at the cost of a 9\% and 3.3\% accuracy drop in vocalized and silent setups, respectively.

For this reason, in the next section, we compare results for two representative configurations: $800\,\mathrm{ms}$, prioritizing responsiveness and ITR, and $1400\,\mathrm{ms}$, prioritizing accuracy.
\vspace{-0.3cm}
\subsection{Incremental Fine Tuning Results}\label{subsec:ft_results}

\begin{figure*}
    \centering
    \includegraphics[width=1.0\linewidth]{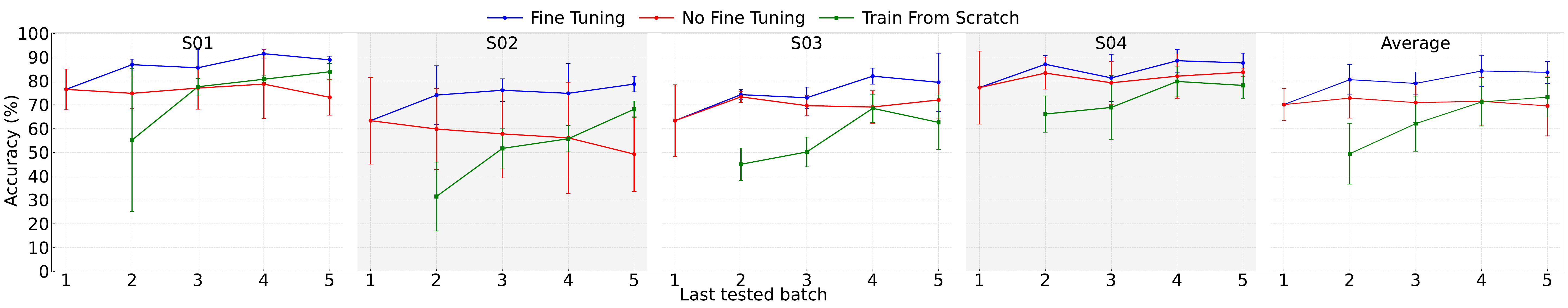}
    \caption{Incremental training performance for vocalized speech (1400\,ms input window). Results are shown per subject (S01–S04) and averaged across subjects (right). The blue curve represents fine-tuning from a model pre-trained on two sessions; the red curve shows the pre-trained model without fine-tuning; the green curve corresponds to training from scratch. For $x=1$, performance is evaluated on the pre-trained model, thus blue and red overlap (no fine-tuning available at this step). Error bars denote standard deviation across sessions (n=3) for individual subjects and across subjects for the average.}
    \label{fig:ft_voc_tmp}
\end{figure*}

\begin{figure*}
    \centering
    \includegraphics[width=1.0\linewidth]{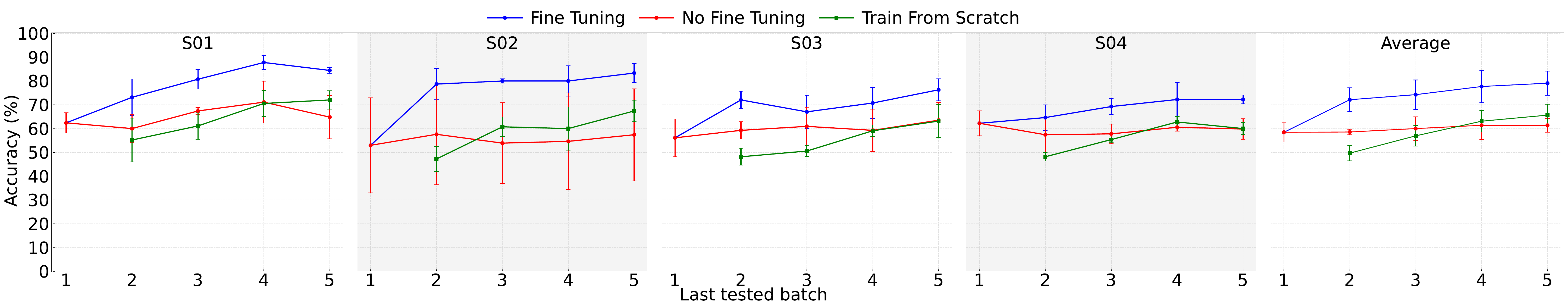}
    \caption{Incremental training performance for vocalized speech (1400\,ms window); methods is equivalent to the vocalized condition.}
    \label{fig:ft_sil_tmp}
    \vspace{-0.5cm}
\end{figure*}

Figures~\ref{fig:ft_voc_tmp} and~\ref{fig:ft_sil_tmp} report the incremental fine-tuning setting results for the vocalized and silent conditions, respectively, using a window length of $1400\,\mathrm{ms}$. Similar trends are observed for the $800\,\mathrm{ms}$ configuration.

In both speaking conditions, fine-tuning consistently improves performance compared to evaluating the model without adaptation (i.e., using the zero-shot scenario with the model pre-trained on two acquisition sessions, without additional fine-tuning epochs, as explained in detail in Sec. \ref{subsec:eval_methods}). Moreover, the fine-tuned models systematically outperform models trained from scratch, confirming the clear benefits of pre-training.

For a window size of $1400\,\mathrm{ms}$, in the vocalized condition, the best average accuracy is achieved after four fine-tuning rounds (84\%). Across the five batches, fine-tuning improves the average accuracy by approximately 8.5\% compared to the non-fine-tuned baseline. 
In the silent condition, peak performance is achieved after five fine-tuning rounds (79\% accuracy), with an average improvement of approximately 12\% across batches compared to the non-fine-tuned model.

Similar behaviour is observed for the $800\,\mathrm{ms}$ window. For vocalized speech, performance peaks after four fine-tuning rounds (78.8\%), reaching an average accuracy of 73.7\%, corresponding to an improvement of approximately 8\% over the non-adapted baseline. 
For silent speech, peak performance is achieved after five fine-tuning rounds (73\% accuracy). The average accuracy across batches reaches 67.0\%, corresponding to an improvement of approximately 10\% compared to the non-fine-tuned model.

From a practical usability perspective, even a single fine-tuning round provides a substantial benefit. With a window size of $1400\,\mathrm{ms}$, one round of adaptation increases accuracy from 70\% to 80\% in the vocalized condition and from 58\% to 72\% in the silent condition. For the $800\,\mathrm{ms}$ configuration, a single fine-tuning round yields an improvement of approximately 7\% in the vocalized case and 12\% in the silent case. In a real-world deployment scenario, this corresponds to recollecting 20 repetitions per word, corresponding to $10\,\mathrm{min}$ of data acquisition. However, this duration could be further reduced through protocol optimizations (e.g., shortening rest intervals between words) and by leveraging a continuous source of ground-truth labeling, such as a microphone, to streamline the annotation process. 




\subsection{Model Deployment}

\begin{table}[t]
\caption{Model deployment on GAP9.}
\label{tab:gap9_performance}
\centering
\begin{tabular}{lcc}
\toprule
\textbf{Input size} & \multicolumn{2}{c}{14 $\times$ 400} \\
\midrule
\textbf{MACs} & \multicolumn{2}{c}{2145984} \\
\midrule
\midrule
\textbf{Frequency [MHz]} & 240 & 370 \\
\textbf{Voltage [V]}     & 0.65 & 0.8 \\
\midrule
Time/inference [ms]        & 2.47  & 1.60  \\
Cycles/inference           & 593497    & 593015    \\
MACs/cycle                 & 3.62    & 3.62    \\
Energy/inference [$\mu$J]  & 63.9 & 97.1 \\
\bottomrule
\end{tabular}
\vspace{-0.5cm}
\end{table}

Table~\ref{tab:gap9_performance} reports the deployment results on the GAP9 microcontroller. We deploy the model trained with an EMG input window of $800\,\mathrm{ms}$. The slight accuracy reduction associated with the shorter window is compensated by a higher \ac{ITR}, as shown in Figs.~\ref{fig:itr_vocalized} and~\ref{fig:itr_silent}. Moreover, inference latency and energy consumption are reduced nearly linearly, resulting in approximately a 50\% reduction.

The \SI{370}{MHz}/\SI{0.8}{V} operating point achieves the lowest latency, reaching \SI{1.60}{ms} for a window size of \SI{0.8}{s}, while the \SI{240}{MHz}/\SI{0.65}{V} configuration minimizes energy per inference, requiring 63.9~$\mu$J. In both operating points, the accelerator sustains $3.62$~MACs/cycle, indicating compute-bound execution on NE16. The network is deployed with 8-bit post-training quantization, resulting in a compact model size of approximately \SI{15.13}{kB}.

For continuous operation, we adopt a sliding-window inference scheme with a step size of \SI{100}{ms}, i.e., one inference every \SI{100}{ms}. Focusing on the most energy-efficient operating point (\SI{240}{MHz}/\SI{0.65}{V}), the total system power is \SI{20.5}{mW}, corresponding to an estimated \SI{27.1}{h} of battery life with a \SI{150}{mAh} battery. The power budget is dominated by EMG acquisition (\SI{15.3}{mW}), followed by data handling and result streaming (\SI{4.55}{mW}), while edge computation contributes only \SI{0.639}{mW}.
\vspace{-0.3cm}
\subsection{Comparison with previous works}
\label{sect:soa-discuss}
\setlength{\lightrulewidth}{0.2pt}   
\begin{table*}[h]
\centering
\scriptsize
\caption{Comparison of EMG-based SSI Systems (System and Application Level)}
\label{tab:ssi_comparison_transposed}
\setlength{\tabcolsep}{4pt}
\renewcommand{\arraystretch}{1.15}

\begin{tabular}{l l c c c c c}
\toprule
\textbf{} & \textbf{} 
& \makecell{Kapur \\ 2018 \cite{kapur_alterego_2018}}
& \makecell{Wang \\ 2021 \cite{wang_2021_all}}
& \makecell{Wu \\ 2025 \cite{wu_towards_2024}}
& \makecell{Tang \\ 2025 \cite{tang_2025_wireless}}
& \textbf{Our Work} \\
\toprule

\multirow{9}{*}{\rotatebox{90}{\textbf{System}}}

& Form Factor 
& \makecell{Plastic Face-Neck\\ Apparatus}
& \makecell{Tattoos +\\ Ear-mounted Electronics}
& Textile Neckband
& Headphone Earmuff
& Textile Neckband \\
\addlinespace[1pt]
\cmidrule(lr){2-7}
& Sensing Location
& Face
& Face
& Neck
& Ears
& Neck \\

\addlinespace[1pt]
\cmidrule(lr){2-7}

& Electrode Type
& \makecell{Wet \\ Gold-Plated \\+ conductive paste$^{a}$}
& Dry Au/Cr on PET
& \makecell{Dry \\ Gold-Plated \\ + Wet Ref.}
& \makecell{Dry\\ Textile-Based \\ Graphene/PEDOT:PSS} 
& \makecell{Dry \cite{datwyler_softpulse} \\ Conductive Polymer-Elastomer \\+ Ag/AgCl coating}\\
\addlinespace[1pt]
\cmidrule(lr){2-7}

& Channels
& 7--CR
& 4--CR
& 10--CR
& 4--CR
& 14--Diff$^{b}$ \\
\addlinespace[1pt]
\cmidrule(lr){2-7}

& Acquisition Electronics
& OpenBCI
& Custom 
& WANDmini \cite{zhou_2019_wandmini}
& Custom
& BioGAP-Ultra \cite{frey_2026_biogapultra}\\
\addlinespace[1pt]
\cmidrule(lr){2-7}
& Connectivity
& BLE
& Bluetooth
& BLE
& WiFi
& BLE \\
\addlinespace[1pt]
\cmidrule(lr){2-7}

& Power
& N.A. 
& N.A.
& 46 mW
& N.A.
& 20.5 mW \\
\addlinespace[1pt]
\cmidrule(lr){2-7}

& Edge AI
& No
& No
& No
& No
& GAP9 \\

\toprule

\multirow{8}{*}{\rotatebox{90}{\textbf{Application}}}

& Vocabulary
& \makecell{10 digits \\ (0-9 numbers)}
& 110 words
& 11 phonemes
& 10 commands
& 8 commands + rest \\
\addlinespace[1pt]
\cmidrule(lr){2-7}
& Speaking Condition
& Silent$^{c}$
& Silent
& Vocalized
& Silent
& Vocalized + Silent \\
\addlinespace[1pt]
\cmidrule(lr){2-7}
& \#Subjects
& 10
& 1
& 2
& 10
& 4 \\
\addlinespace[1pt]
\cmidrule(lr){2-7}
& Dataset Size / Subject
& 750
& 1100
& 110
& 100$^{d}$
& Voc: 2400$^{e}$ ; Sil: 2400$^{e}$ \\
\addlinespace[1pt]
\cmidrule(lr){2-7}
& Repositioning
& No
& No
& No
& No
& Yes \\
\addlinespace[1pt]
\cmidrule(lr){2-7}
& Model
& CNN
& LDA
& Random Forest
& 1D SE-ResNet
& SpeechNet (CNN) \\
\addlinespace[1pt]
\cmidrule(lr){2-7}

& Accuracy (\%)
& 92.0
& 92.6
& 92.7
& 96
& Voc: 84.8$\pm$4.6 \\
& 
& 
& 
& 
& 
& Sil: 77.5$\pm$6.6$^{f}$ \\

\bottomrule
\end{tabular}

\vspace{0.6em}

\begin{flushleft}
\footnotesize
\textit{Diff}: differential; \textit{CR}: common reference;  
\textit{Voc}: vocalized; \textit{Sil}: silent. \\

$^{a}$ Results reported using gold-plated electrodes with conductive paste; $^{b}$ Four channels fully differential; 10 channels differential with electrode sharing; \\ $^{c}$ Authors refer to the speaking condition as internal vocalization; $^{d}$ Data augmentation applied during training; $^{e}$ Effective dataset size is 4800 samples, including rest; $^{f}$ Reported under our \textit{Global Evaluation} setting.

\end{flushleft}
\vspace{-.7cm}
\end{table*}

Table~\ref{tab:ssi_comparison_transposed} compares our solution with previous work.
From a system-level \review{approach}, our solution offers clear advantages in terms of wearability, comfort, and suitability for daily use.
In particular, we propose a minimally obtrusive design based on electrodes embedded in a textile neckband, rather than face-mounted configurations such as those presented by Kapur~\cite{kapur_alterego_2018} and Wang~\cite{wang_2021_all}. 
This design choice aims to improve user acceptance and long-term usability in real-world scenarios. Our design is inspired by the concept proposed by Wu et al.~\cite{wu_towards_2024}. However we adopt a differential acquisition scheme to reject common-mode noise sources. 
This choice eliminates the need for wet reference electrodes, thereby enhancing wearability, ease of use, and long-term comfort.

Powered by the GAP9 platform, our system is, to the best of our knowledge, the first neck-based SSI solution to integrate on-device AI processing capabilities. This represents a key advantage, particularly in privacy-sensitive contexts, as signal processing and inference can be performed locally without transmitting raw biosignals.

During on-device inference, the power consumption is $20.5\,\mathrm{mW}$, enabling continuous operation for $27.1\,\mathrm{h}$ using a $150\,\mathrm{mAh}$ Li-Po battery (2.1$\times$2.8$\times$0.4 cm$^3$, 4g).

From an application point of view, a quantitative comparison across prior SSI studies remains inherently difficult due to the lack of standardized data collection protocols and the absence of uniform evaluation methodologies. Compared to previous works, we intentionally acquired data over multiple days to better reflect realistic usage. This setting introduces additional sources of variability that are known to challenge model generalization: day-to-day physiological changes (e.g., in skin conductivity), domain shifts caused by sensor repositioning between sessions, and potential changes in speaking behavior across days (e.g., articulation pace and consistency). All of these factors can alter EMG amplitude, temporal structure, and overall signal morphology, ultimately increasing the difficulty of the recognition task. In addition, our dataset substantially increases the number of recorded samples relative to most prior work, enabling a more robust assessment. These design choices deliberately raise the difficulty of the problem addressed in this work compared to previous studies.

\section{Conclusion and Limitations}\label{sec:conclusion}

In this work, we extended the results of \cite{meier_2025_parallel}, where we introduced \textit{Silent-Wear}, a truly wearable  SSI based on a textile neckband form factor with on-board signal processing capabilities. 

We presented a new dataset\textsuperscript{\ref{foot:dataset}} collected from four subjects, over three different acquisition sessions spanning three different days, including vocalized and silent EMG recordings of 8 HMI-related commands.

By introducing \textit{SpeechNet}, a tiny (15K parameters) CNN-based architecture for silent and vocalized speech detection, we achieved a cross-validated accuracy of $84.8 \pm 4.6 \%$ for vocalized and $77.5 \pm 6.6 \%$ for silent speech, showcasing superior performance compared to our prior solution \cite{meier_2025_parallel}. 
Furthermore, we conducted an in-depth analysis of the generalization capabilities of our system across multiple days of use, demonstrating an inter-session accuracy of $71.1 \pm 8.3\%$ for vocalized and $59.3 \pm 2.2\%$ for silent speech. To overcome the challenges arising from repositioning, we proposed an incremental fine-tuning strategy, showing the feasibility of more than 10\% accuracy recovery with less than $10\text{min.}$ of data re-collection.
Finally, we deployed our network architecture on the GAP9 \acp{MCU} embedded in BioGAP-Ultra, showcasing the feasibility of end-to-end vocalized and silent speech recognition at the edge, enabling low-latency and private speech detection. Signal acquisition, on-device inference, and wireless transmission of the classification results can be executed with as low as $20.5\,\text{mW}$of power, enabling more than $27.1\,\text{h}$ of continuous use with a $150\,\text{mAh}$ battery.

To ensure widespread use of our solution, promote standardization of data collection methodologies, and enable consistent algorithm performance comparison, we openly release the complete system\footnote{\label{foot:resources}Hardware and firmware: {\scriptsize\url{https://github.com/pulp-bio/BioGAP}} Data collection software: {\scriptsize\url{https://github.com/pulp-bio/biogui/tree/sensors_speech}} \\ Dataset: {\scriptsize\url{https://huggingface.co/datasets/PulpBio/SilentWear}} Code: {\scriptsize\url{https://github.com/pulp-bio/SilentWear}}} (hardware, firmware, software, algorithms and datasets).

Despite the promising results, some limitations remain and open directions for future research. 

First, the current data collection protocol relies on visual trigger cues to prompt articulation. Consequently, subjective reaction time introduces temporal misalignment between trigger onset and the actual speech-related \ac{EMG} activity. This delay may affect segmentation accuracy and partially explains why larger window sizes perform better, as they increase the likelihood of fully capturing the activation pattern. Incorporating a microphone as a continuous ground-truth source would allow automatic alignment of EMG signals with speech onset, reduce reaction-time variability, and streamline the acquisition process. Such an approach would shorten data collection time by removing unnecessary rest intervals and enable larger datasets to be collected within the same time budget. Faster acquisition would also reduce recalibration time during real-world deployment.

Second, although the proposed dataset includes multi-day recordings and realistic repositioning scenarios, it remains limited in terms of the number of subjects and vocabulary size. Expanding the dataset to include a larger and more diverse participant pool, as well as extending beyond isolated commands toward daily-life sentences and continuous speech~\cite{zhou_2025_ave}, would improve validity and enable more comprehensive generalization studies.

Third, the relatively small dataset constrained the exploration of more advanced architectures. With larger-scale data, future work will investigate more advanced models, including pre-trained foundation models based on transformer backbones specifically designed for wearable execution \cite{fasulo_2025_tinymyo}, which have already demonstrated strong performance on speech tasks. 

Fourth, multimodal sensor fusion represents another promising direction. Integrating complementary sensing modalities such as IMU and EEG, which have been explored for speech decoding~\cite{tang_2026_sensing}, may improve robustness to sensor repositioning and physiological variability, ultimately increasing system reliability.

Finally, online system validation remains to be investigated. Evaluating the system under real-time, closed-loop conditions with user interaction and adaptive recalibration will be essential to assess long-term robustness, usability, and readiness for widespread use.

\vspace{-0.5cm}
\section{Acknowledgment}
The authors thank Mattia Orlandi (University of Bologna) for technical support.

The authors used ChatGPT (OpenAI, GPT-4) for language editing, formatting assistance, code proofreading, and the generation of part of the abstract illustrative figure. All scientific concepts, experimental results, and conclusions were independently developed and validated by the authors.
\bibliographystyle{ieeetr}
\bibliography{bib}
\end{document}